\documentclass[conference]{IEEEtran}
\usepackage{graphicx}
\usepackage{tabularx}

\usepackage[T1]{fontenc}
\usepackage[utf8]{inputenc}
\usepackage[english]{babel}

\usepackage{csquotes}
\usepackage{nicefrac}
\usepackage{afterpage}
\usepackage{needspace}
\usepackage[textsize=scriptsize]{todonotes}
\usepackage{booktabs}
\usepackage{tikz}
\usepackage{quantikz}
\usetikzlibrary{arrows,positioning} 
\usetikzlibrary{shapes.geometric, positioning}
\usetikzlibrary{fit, quotes}
\usetikzlibrary{shapes.geometric, positioning}
\usetikzlibrary{automata,arrows}
\usepackage{flushend}
\usepackage{hyperref}
\usepackage{soul}
\usepackage{amssymb}
\usepackage{csvsimple}
\usepackage{amsthm}
\usepackage{lipsum}
\usepackage{varwidth}
\usepackage{utfsym}
\usepackage{fontawesome5}
\usepackage{multicol}
\usepackage{listings}
\usepackage{adjustbox}
\usepackage{float}
\usepackage{placeins}
\usepackage{yquant}
\usepackage{pgfplots}
\usepackage[detect-all=true]{siunitx}
\usepackage{etoolbox}
\usepackage{bbm}
\usepackage{balance}
\usepackage{changes}
\usepackage{makecell}
\usepackage{multirow}

\robustify\bfseries

\makeatletter
\let\MYcaption\@makecaption
\makeatother
\usepackage[font=footnotesize]{subcaption}
\makeatletter
\let\@makecaption\MYcaption
\makeatother

\newtheorem{example}{Example}

\usepackage[backend=biber,
    style=ieee,
    citestyle=numeric-comp,
    sortlocale=en_US,
    sortcites=true,
    url=true,
    eprint=true,
    giveninits=false,
    dashed=false,
    minnames=1,
    maxnames=5]{biblatex}
\AtEveryBibitem{
  \clearname{editor}
  \clearfield{series}
  \clearfield{isbn}
  \clearfield{issn}
  \clearfield{volume}
  \clearfield{number}
  \clearfield{pages}
  \clearfield{doi}
}

\newif\ifdoubleblind

\makeatletter
\newcommand{\redacted}[2][]{
  \ifdoubleblind
    \if\relax\detokenize{#1}\relax
      [redacted for double-blind review]%
    \else
      #1%
    \fi
  \else
    #2%
  \fi
}

\newcommand{\redactedHyper}[1]{
  \ifdoubleblind \else#1\fi
}

\newcommand{\nothing}{}
\makeatother

\definecolor{darkgreen}{RGB}{0, 150, 0}
\lstdefinelanguage{mqt-qasm}{
  keywords={include, qreg, creg, gate, measure, reset, barrier, if, opaque, bit, qubit},
  keywordstyle=\color{blue}\bfseries,
  ndkeywords={x, h, cccx, ccx, cx, z, oracle, diffusion, cp, p},
  ndkeywordstyle=\color{darkgray}\bfseries,
  identifierstyle=\color{black},
  sensitive=false,
  comment=[l]{//},
  morecomment=[s]{/*}{*/},
  commentstyle=\color{darkgreen}\ttfamily,
  stringstyle=\color{red}\ttfamily,
  morestring=[b]",
  morestring=[b]',
  basicstyle=\ttfamily\footnotesize,
  numbers=left,
  numberstyle=\tiny\color{gray},
  stepnumber=1,
  numbersep=5pt,
  showspaces=false,
  showstringspaces=false,
  showtabs=false,
  tabsize=2,
  breaklines=true,
  breakatwhitespace=false,
  captionpos=b,
  frame=single,
  alsoletter=-,
  lineskip=0pt,
}

\lstdefinestyle{qasm}{
    commentstyle=\color{darkgreen},
    keywordstyle=\color{blue},
    numberstyle=\tiny\color{gray},
    stringstyle=\color{red},
    basicstyle=\ttfamily\footnotesize,
    breakatwhitespace=false,         %
    breaklines=true,                 %
    captionpos=t,                    %
    keepspaces=true,                 %
    numbers=left,                    %
    numbersep=5pt,                   %
    showspaces=false,                %
    showstringspaces=false,          %
    showtabs=false,                  %
    tabsize=2,                       %
    frame=single,                    %
}
\newfloat{lstfloat}{htbp}{lop}
\floatname{lstfloat}{\footnotesize Listing}

\redacted[\nothing]{\usepackage{microtype}}

\doubleblindfalse
\renewcommand{\baselinestretch}{0.87}

\hypersetup{
	pdftitle={Qubit Reuse Beyond Reorder and Reset},                      %
	pdfsubject={},                    %
	pdfauthor={\redactedHyper{Damian Rovara, Lukas Burgholzer, Robert Wille}}    %
}

\begin{document}

\renewcommand*{\figureautorefname}{Fig.}
\renewcommand*{\sectionautorefname}{Section}
\renewcommand*{\subsectionautorefname}{Section}
\def\exampleautorefname{Example}

\title{Qubit Reuse Beyond Reorder and Reset\\\vspace{0.2em}\Large Optimizing Quantum Circuits by Fully Utilizing the Potential of Dynamic Circuits\redacted[\vspace*{-3em}]{\vspace*{-0.4cm}}}

\author{
  \redacted[\nothing]{
\IEEEauthorblockN{Damian Rovara\IEEEauthorrefmark{1}\hspace*{1.5cm}Lukas Burgholzer\IEEEauthorrefmark{1}\IEEEauthorrefmark{2}\hspace*{1.5cm}Robert Wille\IEEEauthorrefmark{1}\IEEEauthorrefmark{3}\IEEEauthorrefmark{2}}
\IEEEauthorblockA{\IEEEauthorrefmark{1}Chair for Design Automation, Technical University of Munich, Germany}
\IEEEauthorblockA{\IEEEauthorrefmark{2}Munich Quantum Software Company GmbH, Garching near Munich, Germany}
\IEEEauthorblockA{\IEEEauthorrefmark{3}Software Competence Center Hagenberg GmbH (SCCH), Austria}
\IEEEauthorblockA{\href{mailto:damian.rovara@tum.de}{damian.rovara@tum.de}\hspace*{1.5cm}\href{mailto:lukas.burgholzer@tum.de}{lukas.burgholzer@tum.de}\hspace*{1.5cm}\href{mailto:robert.wille@tum.de}{robert.wille@tum.de}\\
\url{https://www.cda.cit.tum.de/research/quantum}}
	\vspace*{-3em}
  }
}

\maketitle

\begin{abstract}
Qubit reuse offers a promising way to reduce the hardware demands of quantum circuits, but current approaches are largely restricted to reordering measurements and applying qubit resets.
In this work, we present an approach to further optimize quantum circuits by fully utilizing the potential of dynamic quantum circuits—more precisely by moving measurements and introducing dynamic circuit primitives such as classically controlled gates in a way that forges entirely new pathways for qubit reuse.
This significantly reduces the number of required qubits for a variety of circuits, creating new opportunities for running complex circuits on near-term devices with limited qubit counts.
We show that the proposed approach drastically outperforms existing methods, reducing qubit requirements where previous approaches are unable to do so for popular quantum circuits such as \emph{Quantum Phase Estimation (QPE)}, \emph{Quantum Fourier Transform~(QFT)}, and \emph{Variational Quantum Eigensolver (VQE) ansätze}, as well as leading to improvements of up to $95\%$ for sparse random circuits.
\end{abstract}

\section{Introduction}
\label{sec:introduction}

While state-of-the-art quantum computers keep improving in terms of gate fidelity and the number of available qubits, quantum resources are still far from being abundantly available.
As a result, accounting for the number of qubits allowed by a device plays a substantial role in the development of quantum circuits and aiming to reduce it as much as possible remains crucial.
At the same time, the limited degree of connectivity on a large number of quantum computing modalities, such as superconducting quantum computers, also restricts the potential for interactions among individual qubits in quantum circuits.
To execute circuits that require more sophisticated connectivities among qubits, further gates, such as SWAP gates, must be introduced during the mapping procedure~\cite{holmes2020, wille2023,zulehner2019, siraichi2018, fosel2021, wagner2023, li2019, guo2025}, causing the circuit fidelity to fall due to the increased potential of gate errors.

These challenges have spurred significant research into circuit compilation methods designed to mitigate their negative impact~\cite{quetschlich2025mqt, wille2023,zulehner2019, siraichi2018, fosel2021, wagner2023, li2019, guo2025,decross2023, paler2016, brandhofer2023, ding2020, hua2023, pawar2024, sadeghi2022}.
Several approaches have been proposed to reduce the gate overhead during the mapping process and eliminate unnecessary gates to keep the fidelity high.
With the advancement of \emph{dynamic quantum circuits}, multiple methods have similarly been discovered to reduce the number of required qubits for the execution of quantum circuits where possible~\cite{decross2023, paler2016, brandhofer2023, ding2020, hua2023, pawar2024, sadeghi2022}.
This is achieved by reusing qubits that are no longer needed rather than allocating new qubits.
By selecting an optimal order of performing measurements, this reuse of qubits can be taken advantage of repeatedly in various different circuits.
Not only do such approaches help circumvent the limited qubit counts of modern quantum computers, they also provide ways to simplify the required mapping overhead~\cite{brandhofer2023}.

However, these methods for \emph{qubit reuse} are still severely limited:
For circuits with a high degree of interaction between qubits, applying qubit reuse is difficult and often even impossible, as the tightly coupled qubits need to remain accessible at the same time.
While this makes existing qubit reuse methods unable to provide any form of improvements, dynamic quantum circuits provide opportunities to reduce this degree of interaction that has so far remained untapped.
More precisely, by moving measurements through the circuit and then replacing controlled gates with known measurement outcomes, a large number of these interactions can be eliminated, possibly lifting the blockades for qubit reuse.

In this work, we propose methods to fully utilize this potential of dynamic quantum circuits, i.e., to commute measurements through quantum circuits and then use their outcomes to eliminate two-qubit gates.
This allows qubit reuse to be employed in a variety of cases where existing methods only achieve limited results or no improvements at all.
Evaluations show considerable reductions in the number of required qubits for a variety of quantum circuits, outperforming the state of the art for several popular types of quantum circuits.
More precisely, for \emph{Quantum Phase Estimation (QPE)} and \emph{Quantum Fourier Transform (QFT)}, it achieves a reduction to two or one qubits, respectively, while the state-of-the-art method is unable to apply any amount of qubit reuse for these circuits.
Some instances of \emph{hardware-efficient Variational Quantum Eigensolver (VQE) ansätze} are similarly reduced to constant sizes, even in cases when the \mbox{state-of-the-art} approach is unable to perform any qubit reuse.
Furthermore, for random circuits, the proposed approach is able to reduce the total number of qubits by up to 95\% in some cases while requiring a negligible runtime even for deeper circuits consisting of dozens of qubits.

The remainder of this work is structured as follows: \autoref{sec:background} provides an overview on dynamic quantum circuits and illustrates the idea of using them to reduce qubit requirements on a running example.
Then, \autoref{sec:motivation} goes into further detail on existing methods, highlighting the missed potential of dynamic quantum circuits and proposing a general idea to address them.
\autoref{sec:methods} then defines the proposed methods in more detail while \autoref{sec:evaluation} summarizes the results of a conducted thorough evaluation, comparing them with the state of the art for a variety of different circuits.
Finally, \autoref{sec:conclusion} concludes this work.

\section{Background}
\label{sec:background}

This section establishes the basic concepts behind dynamic quantum circuits and then illustrates the trade-off between qubit count and circuit depth by showing how quantum wires can be reduced for multiple computations.
To this end, we use the example of the \emph{Quantum Phase Estimation algorithm}, which also serves as a running example throughout the remainder of this work.

\subsection{Dynamic Quantum Circuits}

Due to the limitations of early-day quantum computers, a large number of quantum circuits are developed using only unitary operators, with measurements located at the very end of execution.
While this paradigm of quantum circuits provides a variety of different applications, recent hardware advancements have increasingly enabled the implementation of further circuit primitives that go beyond unitary operations, such as mid-circuit measurements, classically controlled gates, and reset operations~\cite{corcoles2021, pino2021, egger2018, mcclure2016}.

These components allow the full exploitation of classical computing power during the execution of quantum circuits.
Through \emph{mid-circuit measurements}, the state of individual qubits can be extracted during execution and \emph{classically controlled gates} allow developers to use these measurement results, as well as other computations, to determine what quantum gates should be applied.
Finally, \emph{reset operations} can be used to return qubits to the $\ket{0}$-state, so that they can be dynamically reused in later parts of the execution, rather than requiring new qubits to be allocated.
Not only do these classical circuit primitives provide new ways to control quantum computers, playing a substantial role, for instance, in the implementation of \emph{fault-tolerant quantum circuits}~\cite{gottesman1997stabilizer}, but they also allow for substantial reductions in the resources required for the execution of certain circuits.

The availability of mid-circuit measurements also opens up the consideration of finding suitable locations in a circuit to perform measurements.
In this context, the \emph{deferred measurement principle}~\cite{nielsen2010} states that measurements within a circuit can also be pushed back to a later point without changing the measurement outcomes.
In particular, the measurements can even be pushed through controlled gates that use the measured qubit as a control without affecting the correctness of the applied gate.
This allows even more freedom in the development of dynamic circuits.

\subsection{Reducing Qubit Count Using Dynamic Quantum Circuits}
\label{sec:reducing}
\begin{figure}
    \centering
    \scalebox{0.5}{
        \begin{quantikz}[column sep=0.2cm]
            \lstick{$|\psi\rangle$} & \qw      & \gate{U}           & \gate{U^2} & \gate{U^4}  & \qw      & \qw                      & \qw      & \qw                      & \qw                      & \qw      & \qw      & \qw & \qw & \qw  \\
            \lstick{$|0\rangle$}    & \gate{H} & \ctrl{-1}                 & \qw                      & \qw                       & \qw      & \qw                      & \qw      & \gate{P(-\frac{\pi}{2})} & \gate{P(-\frac{\pi}{4})} & \gate{H} & \meter{} & \qw & \qw & \qw\\
            \lstick{$|0\rangle$}    & \gate{H} & \qw                      & \ctrl{-2}                 & \qw                       & \qw      & \gate{P(-\frac{\pi}{2})} & \gate{H} & \ctrl{-1}                 & \qw                      & \qw      & \qw & \meter{} & \qw & \qw\\
            \lstick{$|0\rangle$}    & \gate{H} & \qw                      & \qw                      & \ctrl{-3}                  & \gate{H} & \ctrl{-1}                 &          & \qw                      & \ctrl{-2}                 & \qw      & \qw & \qw & \meter{} & \qw\\
            \setwiretype{c} \lstick{$c$} &  \qwbundle{3} \cw  & \cw                       & \cw                     & \cw      & \cw  & \cw         & \cw      & \cw       & \cw                                   &  \cw &  \arrow[r,draw=none,"0", pos=-0.15,yshift=0.2cm] \vcw{-3} & \arrow[r,draw=none,"1", pos=-0.15,yshift=0.2cm] \vcw{-2} & \arrow[r,draw=none,"2", pos=-0.25,yshift=0.2cm] \vcw{-1} & \cw \\
        \end{quantikz}
    }\vspace*{-3mm}
    \caption{Quantum phase estimation using three counting qubits.}\label{fig:qpe}\vspace{-3mm}
    \vspace{-1em}
\end{figure}
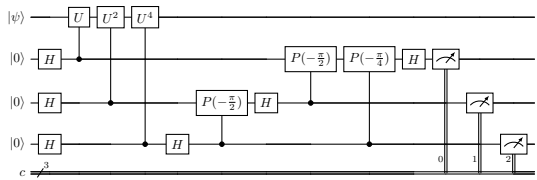

To show how dynamic circuits can be taken advantage of to reduce the number of required qubits, we consider a commonly encountered example, which is used as a running example for the remainder of this work.
\emph{Quantum Phase Estimation}~(QPE,~\cite{nielsen2010}) is a well-known algorithm in quantum computing and a core component of Shor's algorithm for integer factorization~\cite{shor1999}.
\autoref{fig:qpe} shows an implementation of QPE with one eigenstate qubit $\ket{\psi}$ and three counting qubits starting in the state $\ket{000}$.
However, the properties of this algorithm make its execution on state-of-the-art quantum computers challenging.
Due to the limited connectivity, many modern devices cannot efficiently support the all-to-all connectivity requirements of QPE.
This introduces a significant overhead when preparing circuits for execution on such devices, as further gates must be added to circumvent these limitations, drastically impacting runtime and fidelity~\cite{holmes2020, wille2023,zulehner2019, siraichi2018, fosel2021, wagner2023, li2019, guo2025}.

\begin{figure}
    \centering
    \hspace{-0.5cm}
    \scalebox{0.5}{
        \begin{quantikz}[column sep=0.2cm]
            \lstick{$|\psi\rangle$}                     & \qw      & \gate{U^4}  & \qw      & \qw                                                      & \qw                                   & \qw      & \gate{U^2}  & \qw                                 & \qw      & \qw      & \qw                                   & \qw      & \gate{U}  & \qw                                 & \qw                                 & \qw      & \qw      & \qw \\
            \lstick{$|0\rangle$}                             & \gate{H} & \ctrl{-1}        & \gate{H} & \meter{}                                                 & \ground{} \hspace{0.2cm} \ket{0} \hspace{0.2cm} & \gate{H} & \ctrl{-1}        & \gate{P\left(-\frac{\pi}{2}\right)} & \gate{H} & \meter{} & \ground{} \hspace{0.2cm} \ket{0} \hspace{0.2cm} & \gate{H} & \ctrl{-1}       & \gate{P\left(-\frac{\pi}{2}\right)} & \gate{P\left(-\frac{\pi}{4}\right)} & \gate{H} & \meter{} & \qw \\
            \setwiretype{c} \lstick{$c$} &  \qwbundle{3} \cw           & \cw              & \cw      & \arrow[r,draw=none,"2", pos=-0.15,yshift=0.2cm] \vcw{-1} & \cw                                   & \cw      & \cw              & \ctrl[vertical wire=c, wire style={"2"}]{-1}          & \cw      & \arrow[r,draw=none,"1", pos=-0.15,yshift=0.2cm] \vcw{-1} & \cw                                   & \cw      & \cw             & \ctrl[vertical wire=c, wire style={"1"}]{-1}          & \ctrl[vertical wire=c, wire style={"2"}]{-1}          & \cw      & \arrow[r,draw=none,"0", pos=-0.25,yshift=0.2cm] \vcw{-1} & \cw \\
        \end{quantikz}
    }\vspace*{-4mm}
    \caption{Iterative quantum phase estimation using a single counting qubit.}\vspace*{-5mm}
    \label{fig:iqpe}
    \vspace{-0.5em}
\end{figure}
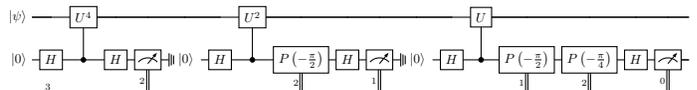

To address this issue, an alternative implementation of the algorithm, also known as \emph{Iterative Quantum Phase Estimation}~(IQPE,~\cite{dobsicek2007}) has been proposed.
Rather than preparing and measuring all counting qubits simultaneously, this approach (illustrated in \autoref{fig:iqpe}) operates on a single counting qubit at a time over multiple iterations.
By employing \mbox{mid-circuit} measurements, classically controlled gates, and reset operations, it reduces the required number of counting qubits to just a single one.
The two approaches can be shown to be equivalent by replacing the reset operations in IQPE with new qubits and employing the \emph{deferred measurement principle}~\cite{nielsen2010} to transform it back to the original QPE version~\cite{burgholzer2022}.

While transforming QPE to IQPE generally increases the circuit depth, the smaller number of required qubits allows IQPE to be executed even on comparatively small quantum computers.
Furthermore, as a two-qubit quantum circuit can be mapped to any desired topology without introducing overhead in the number of gates, IQPE might still outperform the non-iterative approach in terms of actual circuit depth.
As a result, IQPE can achieve better results than QPE on such devices, as demonstrated in~\cite{corcoles2021, mohammadbagherpoor2019}.

\section{Motivation}
\label{sec:motivation}

As shown in \autoref{sec:background}, the introduction of dynamic circuit primitives, such as mid-circuit measurements, resets, and classically controlled gates, can drastically improve the performance of quantum circuits.
However, the transformation from QPE to IQPE was based solely on manual reformulations of the algorithm.
To make this approach more generally applicable, automated methods for qubit reuse are required.

\subsection{Related Work}\label{sec:related}

Existing strategies for this approach are already available, providing ways to automatically select an order in which measurements should be performed, such that qubit reuse can lead to substantial reductions in resource requirements~\cite{decross2023, paler2016, brandhofer2023, ding2020, hua2023, pawar2024, sadeghi2022}.

In 2016, Paler et al.~\cite{paler2016} presented initial methods for qubit reuse that reduce the required number of resources by more than 90\% for certain circuits.
In 2022, Hua et al.~\cite{hua2023} proposed further methods, investigating the trade-off between resource requirements, fidelity, gate count, and circuit duration.
Further, Brandhofer et al.~\cite{brandhofer2023} have presented an algorithm for qubit reuse that takes target device topologies into account.
This allows for a reduced number of swap gates when mapping the circuit to a target architecture, increasing the expected fidelity of the circuit for near-term quantum devices.

In 2023, DeCross et al.~\cite{decross2023} proposed several algorithms to compute such a measurement reordering, including an exact method using constraint programming and an approximation using a local greedy heuristic.
Through an experimental evaluation on several QAOA circuits of varying sizes, they have shown substantial improvements even on large circuits.

However, a common limitation of all of these approaches is that, while they actively try to reduce resource requirements by using mid-circuit measurements and resets, they do not use the full potential of capabilities offered by dynamic quantum circuits.
More precisely, they do not utilize classically controlled gates to broaden the applicability of their algorithms.
As a result, none of the approaches reviewed above would be able to transform QPE to IQPE as described in \autoref{sec:background}.
That is, while existing methods already achieve considerable reductions in the required number of qubits, there is still further potential through the full exploitation of the capabilities of dynamic quantum circuits.

\subsection{General Idea}

To fully utilize the potential of dynamic quantum circuits, we propose a method to trade quantum operations before a measurement for classical operations after the measurement.
This can result in the elimination of gates that are blocking qubit reuse and, therefore, create new possibilities to apply existing algorithms.

The goal is to move measurements before as many gates as possible, which is only possible if the effects of the gates can be applied classically.
For instance, moving a measurement over a Pauli $X$ gate requires the measurement outcome to be flipped.
If this is not possible (such as when applying a Hadamard gate), the measurement cannot be moved any further and has to remain at its current position.%

When a controlled gate is directly followed by a measurement, the measurement can be moved before the control qubit as the deferred measurement principle states that measurements before or after controlled gates are equivalent if they target a control qubit.
As part of that process, the controlled gate can be replaced by a classically controlled gate using the measurement outcome instead of the control qubit.
As this eliminates the interaction of the gate's control and target qubits, it may increase the chances for qubit reuse by reducing their mutual dependencies.
The idea is illustrated using the running example:

\begin{example}
\label{ex:qpe-reuse}
Consider again the QPE circuit in \autoref{fig:qpe}.
Notice that the measurements of the bottom two counting qubits directly follow controlled gates where the qubits are only used as controls.
Therefore, the measurements of those qubits can be moved before the gates.
For the bottom-most qubit, in particular, the measurement can be further moved before the previous controlled phase gate as well.
As these measurements now give us Boolean values for the states of the qubits, we can use the measurement outcomes to replace these controlled phase gates by classically controlled gates. 
As the counting qubits now no longer require overlapping lifetimes, the required resources can be reduced by only using a single qubit that is reset after each individual computation is completed, rather than a three-qubit counting register.
By reordering the instructions in the original program, we can therefore create a new implementation of QPE that only requires a total of two qubits.
This results in the implementation of IQPE as shown in \autoref{fig:iqpe}.
\end{example}

As the example has shown, this method allows QPE to be fully transformed into IQPE.
Furthermore, as shown later in~\autoref{sec:evaluation}, it opens up a vast new range of applications for qubit reuse in which otherwise no reuse can be employed at all,
leading to considerable reductions in resource requirements exceeding previous results in qubit reuse.

\section{Enhanced Qubit Reuse by\\Moving Measurements}
\label{sec:methods}

This section describes an approach that uses the potential motivated above.
First, we propose a novel set of commutation rules that enable the movement of measurements and transformation rules to restructure dynamic quantum circuits.
Based on that, we then describe how to use these rules to facilitate enhanced qubit reuse and reduce resource requirements.

\subsection{Measurement Commutation Rules}
\label{sec:commutation}

To enable the potential for qubit reuse described above, we aim to move measurements as far towards the start of the circuit as possible.
Using a set of commutation rules, we determine whether a measurement can be swapped with the gate preceding it.
If so, we move the measurement before the gate, applying any additional operations required by the commutation rule.
We then repeat this process until the measurement can no longer be moved.
The following rules are proposed:

\paragraph*{Diagonal Gates} Diagonal gates do not influence the outcome of a measurement.
Therefore, any such gates commute with a measurement without requiring any further changes. 
In particular, this applies to all phase gates such as the Pauli $Z$ gate or controlled $P$ gates.

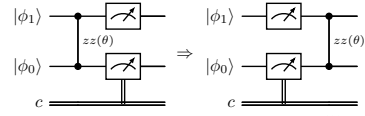
\begin{center}
    \vspace{-0.5em}
    \scalebox{0.65}{
    \begin{quantikz}
        \lstick{$|\phi_1\rangle$} & \ctrl[wire style={"zz(\theta)"}]{1} & \meter{} & \qw  \\
        \lstick{$|\phi_0\rangle$} & \ctrl{-1} & \meter{} & \qw  \\
        \setwiretype{c} \lstick{$c$}         & \cw   & \cw \vcw{-1}  & \cw 
    \end{quantikz}
    $\Rightarrow$
    \begin{quantikz}
        \lstick{$|\phi_1\rangle$} & \meter{} & \ctrl[wire style={"zz(\theta)"}]{1} & \qw  \\
        \lstick{$|\phi_0\rangle$} & \meter{} & \ctrl{-1} & \qw  \\
        \setwiretype{c}\lstick{$c$}         & \cw \vcw{-1}  & \cw & \cw
    \end{quantikz}}
    \vspace{-0.5em}
\end{center}

\paragraph*{Bit-Flip Gates} Gates that cause a flip of the qubit's state can be simulated by negating the classical measurement outcome.
This is because the flip of a qubit's state before the measurement would cause the measurement to report the opposite result with equal probability as compared to not applying the gate before the measurement.
This rule can be applied to Pauli $X$ gates, as well as Pauli $Y$ gates by first decomposing them into $Y = iXZ$ and then also applying the rule for phase gates for the remaining $Z$ gate.

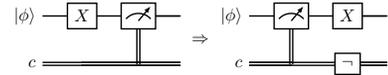
\begin{center}
    \scalebox{0.65}{
    \begin{quantikz}
        \lstick{$|\phi\rangle$} & \gate{X} & \meter{} & \qw  \\
        \setwiretype{c} \lstick{$c$}         & \ghost{\neg}   & \cw \vcw{-1}  & \cw 
    \end{quantikz}
    $\Rightarrow$
    \begin{quantikz}
        \lstick{$|\phi\rangle$} & \meter{} & \gate{X} & \qw  \\
        \setwiretype{c} \lstick{$c$}         & \cw \vcw{-1} &  \gate{\neg} & \cw
    \end{quantikz}}
    \vspace{-0.5em}
\end{center}

\paragraph*{Controlled Gates}
Due to the \emph{deferred measurement principle}, controlled gates can be commuted with measurements on any of their control qubits.
This requires no changes to the classical measurement outcome.

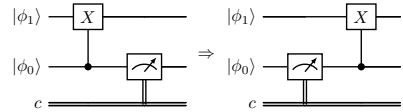
\begin{center}
    \scalebox{0.65}{
    \begin{quantikz}
        \lstick{$|\phi_1\rangle$} & \gate{X} & \qw & \qw  \\
        \lstick{$|\phi_0\rangle$} & \ctrl{-1} & \meter{} & \qw  \\
        \setwiretype{c} \lstick{$c$}         & \cw   & \cw \vcw{-1}  & \cw 
    \end{quantikz}
    $\Rightarrow$
    \begin{quantikz}
        \lstick{$|\phi_1\rangle$} & \qw & \gate{X} & \qw  \\
        \lstick{$|\phi_0\rangle$} & \meter{} & \ctrl{-1} & \qw  \\
        \setwiretype{c} \lstick{$c$}         & \cw \vcw{-1}  & \cw & \cw
    \end{quantikz}}
    \vspace{-0.5em}
\end{center}

These rules can be applied repeatedly, interleaving individual commutation rules to move measurements as far as possible.
The following describes how the resulting circuit is then transformed to reduce the degree of dependencies among qubits to facilitate qubit reuse.

\subsection{Dynamic Circuit Transformations}

After moving measurements as discussed above, further transformations may be employed with the goal of removing gates from the remaining circuit.
Each removed two-qubit gate eliminates one interaction between the qubits, making it more likely that qubit reuse can be employed.
These transformations are based on the following principles:

\paragraph*{Dead Gate Elimination} Any gate that does not influence further measurements in the circuit can be removed.
This can be checked using simple circuit analysis: Starting from the corresponding gate in the circuit and continuing towards the circuit end, all encountered gates are marked as successors and the search is propagated from their outputs.
If such a search does not encounter a measurement, the gate is considered unused and can be eliminated.

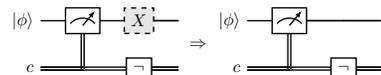
\begin{center}
    \scalebox{0.65}{
    \begin{quantikz}
        \lstick{$|\phi\rangle$} & \meter{} & \gate[style={fill=gray!20, dashed, text=gray}]{X} & \qw  \\
        \setwiretype{c} \lstick{$c$}         & \cw \vcw{-1} &  \gate{\neg} & \cw
    \end{quantikz}
    $\Rightarrow$
    \begin{quantikz}
        \lstick{$|\phi\rangle$} & \meter{} & \qw & \qw  \\
        \setwiretype{c} \lstick{$c$}         & \cw \vcw{-1} &  \gate{\neg} & \cw
    \end{quantikz}}
    \vspace{-0.5em}
\end{center}

\paragraph*{Classical Control Introduction} Qubits that are used as controls directly after a measurement are guaranteed to be in a computational basis state, as the measurement collapses any potential superposition.
As the exact state of the control is known from the classical measurement outcome, the control can be replaced by a classical control.
If the controlled gate uses a negative control rather than a positive one, the classical measurement output must be negated before using it as a classical control for the gate.
Finally, for gates that use multiple controls, only the controls for which measurement outcomes are available can be replaced by those outcomes.
Other controls remain unmodified.

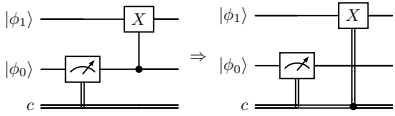
\begin{center}
    \scalebox{0.65}{
    \begin{quantikz}
        \lstick{$|\phi_1\rangle$} & \qw & \gate{X} & \qw  \\
        \lstick{$|\phi_0\rangle$} & \meter{} & \ctrl{-1} & \qw  \\
        \setwiretype{c} \lstick{$c$}         & \cw \vcw{-1}  & \cw & \cw
    \end{quantikz}
    $\Rightarrow$
    \begin{quantikz}
        \lstick{$|\phi_1\rangle$} & \qw & \gate{X} & \qw  \\
        \lstick{$|\phi_0\rangle$} & \meter{} & \qw & \qw  \\
        \setwiretype{c} \lstick{$c$}         & \cw \vcw{-1}  & \ctrl[vertical wire=c]{-2} & \cw
    \end{quantikz}}
    \vspace{-0.5em}
\end{center}

\paragraph*{Exchanging Controls} For controlled phase gates, a target qubit can also be treated as a control qubit.
This is because such gates only affect the state when \emph{all} involved qubits are in the $\ket{1}$ state, so no distinction is made between controls and targets.
Therefore, any such gate, where measurement outcomes are available for the target qubit but not any of the control qubits, can be modified by exchanging the control and target qubits.
This includes gates such as $CZ$ and $CP$.
Afterwards, \emph{classical control introduction} can be employed as described above.

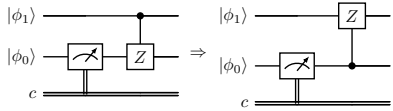
\begin{center}
    \scalebox{0.65}{
    \begin{quantikz}
        \lstick{$|\phi_1\rangle$} & \qw & \ctrl{1} & \qw  \\
        \lstick{$|\phi_0\rangle$} & \meter{} & \gate{Z} & \qw  \\
        \setwiretype{c} \lstick{$c$}         & \cw \vcw{-1}  & \cw & \cw
    \end{quantikz}
    $\Rightarrow$
    \begin{quantikz}
        \lstick{$|\phi_1\rangle$} & \qw & \gate{Z} & \qw  \\
        \lstick{$|\phi_0\rangle$} & \meter{} & \ctrl{-1} & \qw  \\
        \setwiretype{c} \lstick{$c$}         & \cw \vcw{-1}  & \cw & \cw
    \end{quantikz}}
    \vspace{-0.5em}
\end{center}

\subsection{Applying Qubit Reuse}

\newcommand{\oneWireXorGate}{\gate[2][2cm][0.8cm]{}\gateinput[1]{\hspace{-0.15cm}$\begin{array}{l}c_1 \\ c_0\end{array}$} \gateoutput[1]{$\begin{array}{r}c_1 \oplus c_0 \\ c_0\end{array}$\hspace{-0.15cm}}}
\newcommand{\oneWireXorGhost}{\ghost[0.01cm][0.8cm]{}}
\begin{figure*}
    \begin{subfigure}[t]{0.25\textwidth}
        \centering
        \scalebox{0.65}{
        \begin{quantikz}[column sep=0.255cm]
            \lstick{$|\phi_1\rangle$} & \qw & \gate{X} & \qw & \meter{} & \qw  \\
            \lstick{$|\phi_0\rangle$} & \qw & \ctrl{-1} & \meter{} & \qw & \qw \\
            \setwiretype{c} \lstick{$c$}         & \qwbundle{2} \cw  & \cw & \arrow[r,draw=none,"0", pos=-0.15,yshift=0.1cm] \vcw{-1}  & \arrow[r,draw=none,"1", pos=-0.2,yshift=0.1cm] \vcw{-2}  & \oneWireXorGhost
        \end{quantikz}}
        \caption{Starting circuit with two measurements after a $CX$ gate. \vspace{1em}}
        \label{fig:ex-start}
    \end{subfigure}
    \hfill
    \begin{subfigure}[t]{0.3\textwidth}
        \centering
        \scalebox{0.65}{
        \begin{quantikz}[column sep=0.35cm]
            \lstick{$|\phi_1\rangle$}    & \qw & \qw          & \gate{X}  & \meter{}  & \qw  \\
            \lstick{$|\phi_0\rangle$}    & \qw & \meter{}     & \qw & \qw       & \qw \\
            \setwiretype{c} \lstick{$c$} & \qwbundle{2} \cw & \arrow[r,draw=none,"0", pos=-0.2,yshift=0.1cm] \vcw{-1} & \ctrl[vertical wire=c]{-2} \arrow[r,draw=none,"0", pos=-0.2,yshift=0.1cm]      & \arrow[r,draw=none,"1", pos=-0.2,yshift=0.1cm] \vcw{-2}  & \oneWireXorGhost
        \end{quantikz}}
        \caption{Replacing the $CX$ gate by a classically controlled gate after moving the measurement.} 
        \label{fig:ex-a}
    \end{subfigure}
    \hfill
    \begin{subfigure}[t]{0.4\textwidth}
        \centering
        \scalebox{0.65}{
        \begin{quantikz}[column sep=0.35cm]
            \lstick{$|\phi_1\rangle$}    & \qw & \qw          & \meter{}  & \qw & \qw \\
            \lstick{$|\phi_0\rangle$}    & \qw & \meter{}     & \qw       & \qw & \qw \\
            \setwiretype{c} \lstick{$c$} & \qwbundle{2} \cw & \arrow[r,draw=none,"0", pos=-0.15,yshift=0.1cm] \vcw{-1} & \arrow[r,draw=none,"1", pos=-0.2,yshift=0.1cm] \vcw{-2}  & \oneWireXorGate & \cw
        \end{quantikz}}
        \caption{After moving the second measurement through the classically controlled $X$ gate and eliminating dead gates.} 
        \label{fig:ex-b}
    \end{subfigure}
    \begin{subfigure}{\textwidth}
        \centering
        \scalebox{0.65}{
        \begin{quantikz}
            \lstick{$|0\rangle$}    & \qw & \gate[1][0.8cm][0.5cm]{...} & \wire[l][1]["\ket{\phi_0}"{above,pos=0.2}]{a} & \meter{}                                                 & \ground{} \hspace{0.2cm} \ket{0} \hspace{0.2cm} & \gate[1][0.8cm][0.5cm]{...} & \wire[l][1]["\ket{\phi_1}"{above,pos=0.2}]{a} & & \meter{}                                                  & \qw                           & \qw \\
            \setwiretype{c} \lstick{$c$} & \qwbundle{2} \cw & \cw & \cw & \arrow[r,draw=none,"0", pos=-0.15,yshift=0.1cm] \vcw{-1} & \cw                                             & \cw                           & \cw & \arrow[r,draw=none,"1", pos=-0.25,yshift=0.1cm] \vcw{-1}  & \oneWireXorGate & \cw
        \end{quantikz}}
        \caption{Employing qubit reuse to reduce the number of required qubits to 1.} 
        \label{fig:ex-final}
    \end{subfigure}
    \vspace{-1.5em}
    \caption{The process of employing enhanced qubit reuse. Measurements are moved using multiple commutation rules and further transformations eliminate the gates, making the qubits independent and allowing for a reduction in resource requirements.}
    \vspace{-1.5em}
\end{figure*}
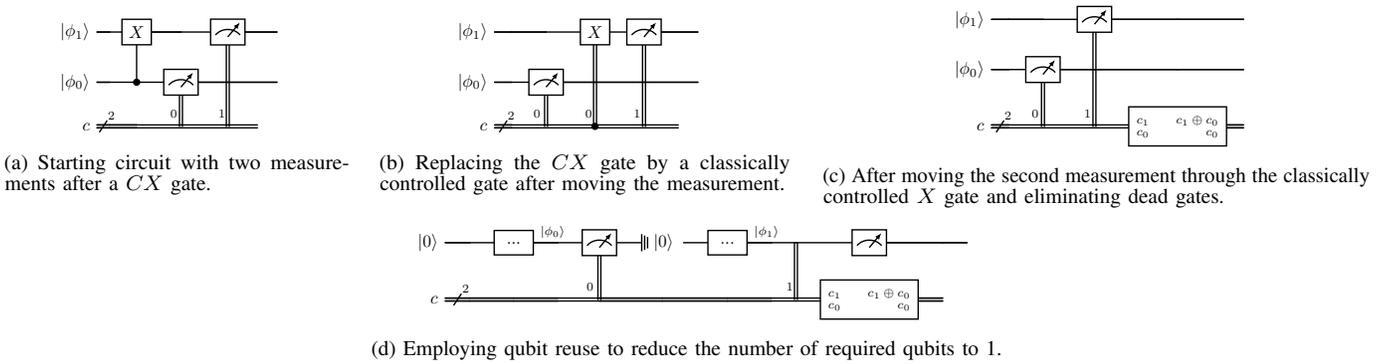

After using the commutation and transformation rules proposed above, the lower number of interactions between qubits may cause qubits that previously had overlapping lifetimes that prevented qubit reuse from being applied to be able to be handled sequentially.
For this work, we implement a direct search to determine the desired measurement order to implement qubit reuse.
To this end, we employ the same circuit analysis as proposed for \emph{dead gate elimination} starting from the first use of qubit $q$.
Whenever a two-qubit gate is encountered, the search is propagated from both outputs of the gate.
If this search reaches the end of the circuit without encountering some other qubit $q'$, then the two qubits are independent and $q$ can reuse $q'$ by resetting $q'$ and appending all gates required for the computation of $q$ to $q'$.
If the search encounters a measurement, we must further ensure that $q'$ does not require the outcome of this measurement for any classically controlled gates.

By repeatedly applying this \emph{search and reset} process, we aim to reduce the required number of qubits as much as possible.
While this approach is not guaranteed to achieve the minimal circuit size as specific reuse orders might converge to a higher number of qubits than others, it allows for a fast execution even as circuits grow, trading solution quality for efficiency.
This guarantees that reuse opportunities can be found even for large circuits that provide the expected use cases for qubit reuse.
However, the proposed algorithm can also be replaced with existing methods by adapting them for the potential classically controlled gates in the circuit.

\begin{example}
    Consider the two-qubit quantum circuit illustrated in \autoref{fig:ex-start}.
    Using the commutation rule for \emph{controlled gates} introduced in \autoref{sec:commutation}, the first measurement can be moved before the $CX$ gate.
    As this results in a measurement directly before a controlled gate, \emph{classical control introduction} can be employed to remove the interaction between the two qubits.
    This results in the circuit shown in \autoref{fig:ex-a}.
    As we now have a measurement directly applied after an $X$ gate, the measurement can be moved before the gate by introducing a negation to the measured classical bit $c_1$.
    However, as the $CX$ gate is only executed if the measurement outcome of~$\ket{\phi_0}$ is equal to $1$, the negation similarly is only performed if $c_0$ is $1$.
    This is equivalent to updating $c_1$ to $c_0 \oplus c_1$.
    As this results in a dead $X$ gate, since no further measurements appear after it, the gate can be removed, resulting in the circuit shown in \autoref{fig:ex-b}.
    Finally, if we assume no other interactions occurred between $\ket{\phi_0}$ and $\ket{\phi_1}$ in previous parts of the circuit, the qubits are \emph{fully independent}.
    Therefore, the computations can be concatenated on a single qubit, resetting the state after measuring $\ket{\phi_0}$, then preparing and measuring $\ket{\phi_1}$.
    The resulting circuit which now only uses a single qubit is shown in \autoref{fig:ex-final}.
\end{example}

\section{Evaluation}
\label{sec:evaluation}

In this section, we evaluate the proposed methods on different use cases.
To this end, we applied the enhanced qubit reuse approach on a wide variety of popular quantum circuits, including \emph{Quantum Phase Estimation} (QPE), \emph{Quantum Fourier Transform} (QFT), and \emph{hardware-efficient Variational Quantum Eigensolver (VQE) ansätze}.
This already provided a detailed insight into the performance of the proposed methods.
To further evaluate this performance for more general circuits, we additionally considered random circuits, focusing on the approach's runtime and result quality as circuit sizes increase.

For all evaluations, we developed an open-source implementation of the proposed approach and made it publicly available at \redacted[]{https://github.com/munich-quantum-toolkit/core}.
This implementation is based on the \emph{Multi-Level Intermediate Representation}~(MLIR,~\redacted[\cite{lattner2021}]{\cite{lattner2021}}) framework, taking advantage of its efficient support for the representation of dynamic quantum circuits and the intuitive methods it provides to implement transformation passes and evaluations over given circuits.
We compare the resulting implementation with the~\texttt{qiskit\_qubit\_reuse} Python package~\cite{qiskit-qubit-reuse} which provides an accessible implementation of the qubit reuse algorithm proposed by \mbox{DeCross}~et~al.~\cite{decross2023}, representing the state of the art in qubit reuse (see \autoref{sec:related}).

\subsection{Evaluation on QPE and QFT Circuits}

To evaluate the applicability of the proposed methods, we first perform evaluations for different sizes of QPE and QFT circuits.
In both cases, the proposed method manages to reduce the number of qubits to the \emph{minimally} possible size.

For QPE, a circuit consisting of $n - 1$ counting qubits and one eigenstate qubit is \emph{always} reduced down to a total of two qubits only, as the eigenstate qubit is required to remain accessible during the execution of the full circuit, while a single counting qubit can be repeatedly reset and recomputed to obtain the phase digits as shown in \autoref{fig:iqpe}.
Furthermore, in certain instances, such as when estimating the phase of a single $P$ gate, additional classically controlled gates are introduced so that the number of gates can even be reduced further than textbook IQPE to \emph{just a single qubit}, eliminating the need for an eigenstate qubit completely.
Also for QFT, the proposed approach manages to compress the $n$-qubit starting circuit down to reduced circuit with \emph{a single qubit only}.

In both cases, the existing state of the art (reviewed in \autoref{sec:related} cannot perform any qubit reuse at all for these circuits.
This shows that the proposed methods provide a much wider range of potential use cases for qubit reuse.
However, as discussed above, this qubit reuse causes a rise in circuit depth, leading to the fundamental trade-off inherent in qubit reuse.

More precisely, \autoref{tab:qpe-qft} shows the corresponding qubit reductions as well as the obtained circuit depths of QPE and QFT after applying the proposed approach for different starting sizes $n$, compared to the original depths before applying qubit reuse.
All reported results shown in \autoref{tab:qpe-qft} were computed in real time, requiring only a maximum of $265$ms and $276$ms for the 50 qubit benchmarks of QPE and QFT, respectively.

The obtained results confirm that the proposed methods can achieve drastic reductions in the number of qubits.
For QPE, starting circuits of \emph{any size} $n$ allow a decrease in the number of required qubits by $n - 2$.
For larger instances, such as the~\mbox{50-qubit} instance considered in this evaluation, this substantial reduction only leads to a depth increase of~$14\%$.
This is because for the application of the controlled $U$ gates, which represents the largest part of the circuit as the number of qubits grows, individual layers in the circuits only consist of a single gate even when using multiple qubits.
Therefore, even when concatenating all layers on a single wire, the depth overhead does not rise significantly.
Similarly, QFT circuits of any size can be reduced to just \emph{a single} qubit.
While depth increases due to qubit reuse are more apparent for QFT, the developer is free to select a desired number of qubits.
This way, the proposed approach allows developers to freely explore and control the trade-off between qubit count and circuit depth.
Additionally, as a reduction in the number of qubits may also simplify the circuit's connectivity constraints, the circuit depth after mapping it to actual quantum devices may once again be improved compared to larger circuits.

Overall, these evaluations confirm that, for important quantum circuits such as QPE and QFT, the proposed approach can automatically reduce the number of qubits from dozens to just two (QPE) or even a single (QFT) qubit---a potential not utilized by existing automated methods.

\begin{table}
    \centering
    \caption{Results obtained from evaluating QPE and QFT circuits.}\vspace*{-3mm}
    \label{tab:qpe-qft}
    \scalebox{0.9}{
    \begin{tabular}{r r r r r r r r}
        \toprule
        \multicolumn{4}{c}{QPE} & \multicolumn{4}{c}{QFT} \\
        \cmidrule(r){1-4}\cmidrule(l){5-8} 
        $n$ & $n_{reused}$ & $d$ & $d_{reused}$ & $n$ & $n_{reused}$ & $d$ & $d_{reused}$ \\ \midrule
         4 & 2 & 9    & 17   &  4 & 1 & 8   & 17 \\
         6 & 2 & 18   & 34   &  6 & 1 & 12  & 32 \\
         8 & 2 & 31   & 55   &  8 & 1 & 16  & 51 \\
        10 & 2 & 48   & 80   & 10 & 1 & 20  & 74 \\
        20 & 2 & 193  & 265  & 20 & 1 & 40  & 249 \\
        30 & 2 & 438  & 550  & 30 & 1 & 60  & 524 \\
        40 & 2 & 783  & 935  & 40 & 1 & 80  & 899 \\
        50 & 2 & 1228 & 1420 & 50 & 1 & 100 & 1374 \\ \bottomrule 
    \end{tabular}}\\\vspace{0.2em}
    {\scriptsize $n$: Number of qubits \hspace*{0.4cm} $d$: Circuit depth}
    \vspace{-2em}
\end{table}

\subsection{Evaluation on VQE Circuits}

\begin{table}
    \centering
    \caption{Results obtained from evaluating hardware efficient VQE ansatz circuits.}
    \label{tab:vqe}
    \begin{tabular}{r r r r r}
    \toprule
    \multirow{2}{*}[-0.5ex]{\makecell{Entanglement\\Strategy}} & \multicolumn{2}{c}{State-of-the-Art} & \multicolumn{2}{c}{Proposed Approach}\\
    \cmidrule(lr){2-3} \cmidrule(l){4-5}
     & $n_{reused}$ & $d_n$ & $n_{reused}$ & $d_n$ \\ \midrule
    circular        & 3 & 2.39 & 2 & 3.39 \\
    pairwise        & 4 & 1.39 & 2 & 2.30 \\
    linear          & 2 & 2.58 & 1 & 3.79 \\
    reverse-linear  & 2 & 2.51 & 2 & 3.39 \\
    full            & $n$ & 1.76 & 1 & 7.38 \\ \bottomrule
    \end{tabular}\\ \vspace{0.2em}
    {$n_{reused}$: Number of qubits after applying qubit reuse \\ $d_n$: Depth of the resulting circuit, normalized by number of qubits}
    \vspace{-2.5em}
\end{table}

\begin{table*}[!t]
    \centering
    \caption{Results obtained from evaluating random circuits.}
    \label{tab:random}
    \begin{tabular}{r r r r r r r r r r r r r r}
    \toprule
    \multicolumn{2}{c}{Input} & \multicolumn{4}{c}{State-of-the-Art} & \multicolumn{4}{c}{Proposed Method} & \multicolumn{4}{c}{$\Delta_{\text{state-of-the-art} \rightarrow \text{proposed}}$}\\
    \cmidrule(r){1-2} \cmidrule(l){3-6} \cmidrule(l){7-10} \cmidrule(l){11-14}  
    $n$ & $d$ & $n_{reused}$ & $d_{reused}$ & $|G_2|$ & $t [s]$ & $n_{reused}$ & $d_{reused}$ & $|G_2|$ & $t [s]$ & $n_{reused}$ & $d_{reused}$ & $|G_2|$ & $t$ \\ \midrule
    20 & 2 & 5.2 & 29.6 & 8.6 & 0.0047 & 2.2 & 45.0 & 1.2 & 0.0362 &         57.7\% & -52.0\% & 86.0\% & -670.2\% \\
    20 & 8 & 12.6 & 33.2 & 30.4 & 0.0107 & 7.8 & 56.0 & 19.8 & 0.0486 &      38.1\% & -68.7\% & 34.9\% & -354.2\% \\
    20 & 26 & 19.4 & 30.2 & 98.8 & 0.3937 & 19.2 & 41.4 & 84.4 & 0.0767 &    1.0\% & -37.1\% & 14.6\% & 80.5\% \\
    30 & 2 & 6.8 & 43.8 & 12.6 & 0.0062 & 1.6 & 74.2 & 0.8 & 0.0376 &        76.5\% & -69.4\% & 93.7\% & -506.5\% \\
    30 & 10 & 20.4 & 46.2 & 54.8 & 0.0225 & 12.6 & 90.0 & 35.8 & 0.0519 &    38.2\% & -94.8\% & 34.7\% & -130.7\% \\
    30 & 42 & 30.0 & 40.6 & 239.8 & 82.7132 & 30.0 & 42.8 & 221.4 & 0.0917 & 0.0\% & -5.4\% & 7.7\% & 99.9\% \\
    40 & 3 & 12.0 & 61.8 & 24.0 & 0.0119 & 3.0 & 59.0 & 4.2 & 0.0447 &       75.0\% & 4.5\% & 82.5\% & -275.6\% \\
    40 & 9 & 23.2 & 69.4 & 67.8 & 0.0262 & 14.2 & 109.0 & 45.4 & 0.0623 &    38.8\% & -57.1\% & 33.0\% & -137.8\% \\
    40 & 27 & 39.4 & 35.6 & 209.0 & 1.5640 & 37.6 & 65.8 & 179.0 & 0.1126 &  4.6\% & -84.8\% & 14.4\% & 92.8\% \\
    50 & 3 & 11.6 & 73.0 & 27.8 & 0.0130 & 2.2 & 100.6 & 6.4 & 0.0500 &      81.0\% & -37.8\% & 77.0\% & -284.6\% \\
    50 & 10 & 32.8 & 75.4 & 97.8 & 0.0346 & 18.0 & 125.6 & 59.0 & 0.0779 &   45.1\% & -66.6\% & 39.7\% & -125.1\% \\
    50 & 33 & 49.8 & 34.6 & 314.8 & 13.6188 & 49.4 & 48.0 & 285.2 & 0.1617 & 0.8\% & -38.7\% & 9.4\% & 98.8\% \\ 
    60 & 3 & 15.0 & 87.6 & 36.4 & 0.0111 & 3.4 & 94.0 & 7.8 & 0.0462  & 77.3\% & -7.3\% & 78.6\% & -316.2\%\\
    60 & 12 & 45.4 & 81.4 & 140.6 & 0.0476 & 31.0 & 132.6 & 99.8 & 0.0842  & 31.7\% & -62.9\% & 29.0\% & -76.9\%\\
    60 & 48 & 60.0 & 46.8 & 536.0 & 1134.8826 & 60.0 & 49.0 & 500.8 & 0.1665  & 0.0\% & -4.7\% & 6.6\% & 100.0\%\\\bottomrule%

    \end{tabular}\\ \vspace{0.2em}
    {\scriptsize $n$: Number of qubits \hspace*{0.4cm} $d$: Circuit depth $|G_2|$: Number of two-qubit gates $t$: Runtime of qubit reuse algorithm}
    \vspace{-1.5em}
\end{table*}

To show further applications of the proposed methods for near-term practical examples, we additionally investigated the performance of enhanced qubit reuse for circuits realizing the hardware-efficient \emph{Variational Quantum Eigensolver}~(VQE, ~\cite{peruzzo2014a}) ansatz~\cite{kandala2017hardware}.
This ansatz provides ways for the parametrized preparation of quantum states that are efficiently executable on near-term devices.
It consists of a rotation block and an entanglement block.
For this evaluation, we built rotation blocks from $RX$ gates, while we constructed the entanglement blocks using $CX$ gates.
We considered instances consisting of a single repetition layer, as they provide a proper use case showing the potential of qubit reuse.

The hardware efficient ansatz can be constructed using one of several entanglement strategies that indicate to what degree the qubits interact with each other inside the entanglement block.
As this degree of interaction also affects the potential for qubit reuse, we compared several different entanglement strategies over ansätze of different sizes.
Furthermore, we also investigated to what degree the state-of-the-art implementation can be employed and what improvements are made by this method.

\autoref{tab:vqe} shows the circuit sizes achieved by the proposed appraoch as well as by the \mbox{state-of-the-art} implementation.
It further lists the mean depth of the resulting circuits, divided by the number of qubits in the input circuit for normalization.
Again, all results have been obtained in negligible runtime.

Notably, this evaluation shows that the final number of qubits in the circuit after applying qubit reuse does not depend on the initial circuit size.
Due to the regular structure of these circuits, the repeated use of qubit reuse converges to the same final number of qubits, regardless of the original size $n$.

The results clearly confirm the additional potential of the proposed methods.
While in case of the \emph{reverse-linear} entanglement strategy, both approaches achieve the same results, the proposed method outperforms the state-of-the-art implementation for all other entanglement strategies---further pushing the limits of qubit reuse.
Moreover, in case of the \emph{full} entanglement strategy, the proposed method reduces the number of required qubits to a minimum of one, while the state-of-the-art implementation cannot perform any qubit reuse at all, demonstrating the full potential of the proposed~methods.

These qubit savings, however, come at the cost of increased circuit depths.
However, the results confirm that the increase in depth is often not proportional to the reduction in qubits.
In the end, it is up to the developer to decide on a suitable trade-off between qubit count and circuit depth.
While such a trade-off might not be worth it for small circuits, it becomes crucial for larger circuits that are close to the limits of available hardware.

Overall, this once again confirms the increased potential for qubit reuse offered by the proposed methods.

\subsection{Evaluation of Random Circuits}

While the previous evaluations provided a wide overview on the potential of enhanced qubit reuse for several popular quantum circuits, we additionally conducted further evaluations on random circuits to investigate the approach's scalability in terms of result quality and runtime.
For this, we created several circuits with different qubit counts and depths by randomly concatenating one and two-qubit gates and measuring each qubit at the end of the circuit.

\autoref{tab:random} reports the results achieved by running the proposed appraoch as well as the state-of-the-art implementation on the given circuits.
Each entry shows the mean number of qubits, depth, two-qubit gate count, and execution time, obtained by running the corresponding algorithms on five random circuit instances for each listed number of qubits and circuit depth.
Furthermore, each entry also lists the percentage reduction of each of these statistics obtained by the proposed approach and compared to the state of the art.

The results confirm that, also for these instances, the proposed approach can lead to drastic reductions in the required number of qubits.
However, unlike the previous evaluations, the final number of qubits does depend on the original circuit size, as the random structure of these circuits does not allow for a regular reuse of qubits.
Nonetheless, the proposed approach achieved better results than the state-of-the-art implementation for \emph{all} considered circuit instances.
For low-depth instances, such as $30$ qubits with a depth of $2$, we find a drastic decrease of almost $95\%$, while the existing methods only reduce the qubit requirements by less than $80\%$.
However, as the input depth rises, the differences between the two approaches decreases.
This is because for larger circuits it becomes increasingly unlikely that the proposed methods can eliminate enough \mbox{two-qubit} gates to enable multiple qubits to be reused.

While the depths of the circuits after applying the proposed methods are largely greater than the depths obtained from the existing reference implementation, this may not necessarily lead to a lower fidelity circuit, as discussed in \autoref{sec:reducing}.
This is even more so the case, as the number of two-qubit gates is substantially lower for circuits obtained from the proposed methods and the reduction in qubits is more significant than the increase in circuit depth compared to the state-of-the-art implementation.

Furthermore, the state-of-the-art implementation also puts more effort into obtaining low depths, resulting in substantially higher execution times compared to the proposed approach.
For example, while the execution of the proposed approach took up a maximum of $0.2$ seconds for the largest investigated examples, the state-of-the-art implementation required more than $1\kern0.075em 134$ seconds for the same circuit.
This further highlights the efficiency of the proposed method.
\medskip

Overall, these evaluations showed that the utilization of the additional potential of dynamic quantum circuits as discussed above and employed in the proposed method, can lead to substantial improvements not just for structured circuits such as QPE, QFT, or VQE ansätze, but also find more general applications in arbitrary quantum circuits.
As the proposed method can further be executed in real time, it provides not only a widely but also a cheaply applicable approach to reduce qubit requirements for quantum circuits.

\vspace{-0.5em}
\section{Conclusion}
\label{sec:conclusion}

In this work, we have introduced a novel approach for enhanced qubit reuse.
By moving measurements to earlier parts of the circuit through commutation rules and strategically using them to eliminate two-qubit gates, the proposed approach reduces the interaction between qubits and, in turn, facilitates the application of qubit reuse.
This allows for a substantial reduction in the resource requirements of a variety of quantum circuits where existing methods would otherwise be very limited or not applicable at all.
Through evaluations on several different classes of circuits, we have shown the advantages of the proposed approach, leading to reductions in qubit requirements of up to $95\%$ for sparse circuits and $64\%$ for circuits of medium size and depth.
All components of the proposed approach are made available as part of an open-source implementation at \redacted[]{https://github.com/munich-quantum-toolkit/core}.

\pagebreak

\redacted[\nothing]{
\clearpage
\section*{Acknowledgments}
This work received funding from the European Research Council (ERC) under the
European Union's Horizon 2020 research and innovation program (grant agreement 
No. 101001318), was part of the Munich Quantum Valley, which is supported by 
the Bavarian state government with funds from the Hightech Agenda Bayern Plus, 
and has been supported by the BMK and BMDW.
}

\printbibliography

\end{document}